\documentclass[a4paper,12pt]{iopart}
\usepackage{amssymb}
\usepackage[dvips]{graphicx}
\begin{document}
\title[A.Novikov-Borodin. Phys.Objects of Dark Systems]{Physical Objects of Dark Systems} 
\author{A.Novikov-Borodin\dag\ }
\address{\dag\ Institute for Nuclear Researches of RAS,\\ 60-th October Anniv.pr. 7a, 117312 Moscow, Russia.}
\ead{novikov@inr.ru}
\begin{footnotesize} \hspace{100mm} \today \end{footnotesize} 
%
\begin{abstract} 

The hypothesis of existence of off-site continuums is investigated. Principles of the physical description are formulated. The structure of off-site continuums and opportunities of observation of off-site physical objects from the continuum of the observer is investigated. There are found conformities between properties of the considered objects with properties of known physical objects and fundamental interactions both in micro-, and macro-scales. 

\end{abstract}
%
%
%
\section{\bf Introduction.}

The general theory of relativity in itself generates the conservation laws -- and not as a consequence of field equations, but as identities (E.Schr\"odinger \cite{Schr86}). Only from one fact of the general invariance of the integral: 

\begin{equation}
{\cal I} = \int_{\cal G} \mathfrak{R} \, d^{4} x ,
\label{INT}
\end{equation}  

where $ \mathfrak {R} $ is some scalar density, four identical ratios between Hamiltonian derivatives of density $ {\mathfrak {R}} $ may be got, and these ratios follow such as conservation laws. Integration in (\ref {INT}) is carried out on all ``four-dimensional continuum", thus meaning its continuity and infinity. E.Schr\"odinger \cite {Schr86} indirectly pointed out to the limitation of such approach in physics to the description of the physical reality from the point of view of mathematicians: ``The physicist, however, has got used to consider, that the certain letter ($ \phi $\footnote {a scalar field - Author's comment} in our case) concerns to a {\it concrete} field in {\it any} frame of references. His most important general reasoning concern to ``the general frame of references", which he does not concretize and consequently should not actually change frequently very much, though the principle of invariance relating to transformations is constantly meant". E.Schr\"odinger, however, did not go beyond this remark. What is this ``general frame of references", for whom and how much is it ``general"? Whether there are other ones? And if yes, what consequences can it have from the point of view of physics? 

\section{\bf Space-time continuums.}
\label{sec:Continuum}

In the general theory of relativity the space-time topometry depends on physical objects existing in this space-time (gravitating bodies, particles, fields). ``It is necessary, strictly speaking, to have a set of an infinite number of the bodies filling all space, like some ``medium". Such system of bodies together with connected to each of them arbitrarily clocks is a frame of reference in the general theory of relativity" (L.Landau and E.Lifshitz \cite {LL88}). Such as ``medium" not only fills the space, but also forms it by itself, we do not have bases to deny the existence of other ``medium" -- the space, which can not adjoin to ours, include it in itself or to be itself a part of our ``medium". Arguments, that if we do not observe anything similar to such ``mediums", so it does not exist, or it is no sense to consider, are not convincing, because a question, what could we observe in this case, simply was never investigated. At least, the author does not know anything about such investigations. 

Uniqueness of a continuum is not a necessary condition from the point of view of mathematics. It is declared in the mathematical theory of sets, that, for example, any local area of continuous infinite multidimensional space has the same power of set as all that space in sense that it is possible to establish the point-to-point correspondence between elements of the given sets. It is possible the situation at which the considered continuum is a subset by itself of another continuum, and, on the contrary, can include a number of other continuums. As the elementary example, let's consider the functional correspondences between two four-dimensional continuums $ {\cal G} $: $ \{ x: (x^0, x^1, x^2, x^3) \} $ and $ {\cal G'} $: $ \{ x': (x'^0, x'^1, x'^2, x'^3) \} $ continuous and infinite by each coordinate. So, they have the equal power of set. Transformations of coordinates $x^i = \frac{1}{\pi} \arctan (x'^i) $, $i=0..3 $ transform a four-dimensional continuum $ {\cal G'} $ into a four-dimensional unit cube from $ {\cal G} $: ${\cal G'} \rightleftharpoons D \subset {\cal G}$. Transformations $x'^i = \frac{1}{\pi} \arctan(x^i) $, $i=0..3$ transform a four-dimensional continuum $ {\cal G} $ into a four-dimensional unit cube from $ {\cal G '} $: ${\cal G} \rightleftharpoons D' \subset {\cal G'}$. 

So, the continuum $ {\cal G} $ and scalar density $ \mathfrak {R} $ from (\ref {INT}) generate the ``medium" -- the system including ``a set of infinite number of bodies" and corresponding conservation laws. Other continuum $ \cal G' $ distinguished from $ \cal G $, will generate its own ``medium", i.e. the system including own physical objects, conservation laws and, hence, own space-time structure. This medium will differ, generally speaking, from the medium generated by $ \cal G $. We will call such mediums as {\it systems of physical description}. 

We shall define the continuum $ \cal G' $, conservation laws generated by it and the physical objects in it as {\it off-site} for $ \cal G $. Similarly, $ \cal G $ will be off-site for $ \cal G' $. The set of observers watching their own continuum, we shall name {\it the generalized or system observer}. The generalized observer watching an off-site continuum, will see the {\it seen or observable} parameters of the off-site medium. 

If the off-site continuum $ {\cal G'} $ can be observed in area $ D $ from $ {\cal G} $: $ {\cal G'} \rightleftharpoons D \subset {\cal G} $, so we shall name $ {\cal G '} $ as an {\it enclosed} continuum in relation to $ {\cal G} $, and $ {\cal G} $ will be a {\it containing} one in relation to $ {\cal G'} $. 

There will be some specificity at the description of observable parameters of the enclosed and containing off-site continuums. Anyhow, the off-site enclosed continuums from a continuum containing them can be observed entirely, while only the limited area of an off-site containing continuum is accessible to the observer from the enclosed continuum. We shall define the observation of parameters of the enclosed continuums as an observation in {\it micro-scale}, and of the containing continuums as an observation in {\it macro-scale}. The analysis of the structure of the enclosed continuums will be considered in sections \ref{sec:PhysObj} and \ref{sec:Properties}. The approach to the analysis of the structure of containing continuums will be considered in section \ref{sec:MMScale}.  

\section{\bf Observable structure of the enclosed continuums.}
\label{sec:PhysObj} 

Let's suppose, that some functional conformity $ {\cal F} : {\cal G} \stackrel{\cal F}{\longrightarrow} {\cal G'} $ or $ {\cal F'}: {\cal G'} \stackrel{\cal F'}{\longrightarrow} {\cal G} $ is assigned between elements of two continuums $ {\cal G} $: $ \{x: (x^0, \dots , x^n) \} $ and $ {\cal G'} $: $ \{x': (x'^0, \dots , x'^m) \} $.   

There is no sense to try to analyze all set of possible functional transformations $ {\cal F} $ at once, therefore we shall enter minimally possible physical restrictions. We shall examine an off-site media from the point of view of the generalized observer of ``our" four-dimensional continuum $ {\cal G} $: $ \{x: (x^0, \dots , x^3) \} $, supposing that it is well-known the system of the physical description in it. Particularly, the metrics $ds^2=g_{ik}dx^i dx^k$ is defined in it. We shall consider also, that elements of off-site continuums (or, at least, some subset of their elements) can be organized by the similar way: $ {\cal G '} $: $ \{x': (x'^0, \dots , x'^3) \} $ so, that functional conformity between these elements of continuums can be presented as:  
\begin{eqnarray}
x'^i = f^i (x^0, x^1, x^2, x^3), \qquad &\mbox{if}& \qquad  {\cal G'} \rightleftharpoons D \subset{\cal G}, \\
x^i = f'^i (x'^0, x'^1, x'^2, x'^3),  &\mbox{if}&  \qquad  {\cal G} \rightleftharpoons D' \subset{\cal G'}\nonumber.
\label{function}
\end{eqnarray}  

In the first case the continuum $ {\cal G'} $ is enclosed in relation to $ {\cal G} $, and in the second case it is a containing one. The zero components will be compared to time, and the rest ones -- to spatial components in each of continuums that is necessary for carrying out of the further reasoning within the framework of existing physical concepts.   

The enclosed continuum from the point of view of the observer of the containing system will possess an additional internal structure inside the area of observation $ D $. We shall illustrate it on an example of continuously differentiated up to borders of $D$ functions $ f^i $. In this case their differentials will be transformed according to formulas (as usually, it is meant summation on repeating indexes): $ dx'^i = (\partial f^i /\partial x^k) dx^k = (\partial x'^i /\partial x^k) dx^k. $

For definiteness we shall consider, that in the field of observation $D $ of the systems of the observer $ \cal G $, the Minkowski's metrics $ds^2=dx_i dx^i = (dx^0)^2-dx_{\alpha} dx^{\alpha}, i=0..3, \alpha = 1..3 $ is determined. Let's define the expression $ (ds')^2=dx'_i dx'^i = (dx'^0)^2-dx'_{\alpha} dx'^{\alpha} $ in some local area of the enclosed continuum $ \cal G' $. This expressions will be interpreted by the generalized observer of $ {\cal G} $ as the observable metrics $ (ds^{\cal G'})^2 $ of the continuum $ {\cal G '} $: 

\begin{equation}
(ds^{\cal G'})^2 = \left[ \frac{\partial f^0}{\partial x^i} \frac{\partial f^0}{\partial x^k} - \sum_{\alpha = 1,2,3} \frac{\partial f^{\alpha}}{\partial x^i} \frac{\partial f^{\alpha}}{\partial x^k} \right] dx^i dx^k = g^{{\cal G'}}_{ik} dx^i dx^k . 
\label{metrics}
\end{equation}   

L.Landau and E.Lifshitz \cite {LL88} give the following physical interpretation of types of elements of metric tensor: ``It is necessary to emphasize a difference between sense of a condition $g_ {00}> 0 $ and a condition of the certain signature (signs on principal values) of the tensor $g_{ik} $. The tensor $g _ {ik} $, not satisfying to the second one of these conditions, cannot correspond to any real gravitational field at all, i.e. the metrics of the real space-time. Default of a condition $g_{00}> 0 $ would mean only, that the corresponding frame of references cannot be carried out by real bodies; if the condition on principal values thus is carried out, it is possible to achieve that $g_{00} $ becomes positive by appropriate transformation of coordinates". 

The given interpretations are completely appropriate for physical interpretation of space-time structure of introduced off-site continuums if to take into account, that concepts of ``real space-time" and ``real bodies" are not determined by Landau, and to apply our contents considered in Sect.\ref{sec:Continuum} to these concepts. We shall call this assumption as {\it a principle of observability} of off-site continuums. 

Thus, we consider, that $g^{\cal G'}_{ik} $ (from (\ref{metrics})) is seen by the generalized observer of system $ {\cal G} $ as an observable metric tensor in ${\cal G'} $. The elements of $g^{{\cal G'}}_{ik} $ also will determine the observable physical space-time structure of the system $ {\cal G'} $ and the visual properties of off-site physical objects connected with that ones from $ {\cal G'} $. The observable space-time structure of system $ {\cal G'} $ can be conditionally separated into three areas, hence: the ``timeable" area $g^{\cal G'}_{00} > 0 $; the ``spatial" area $ \det (g^{\cal G'}_{ik}) > 0 $ and the ``transitive" area where both previous conditions are not carried out. The observable time of an off-site continuum in timeable areas can be determined from (\ref{metrics}), assuming $dx^{\alpha} =0, \alpha=1,2,3 $: $d\tau^{\cal G'} = \sqrt {g^{\cal G'}_{00}} dx^0 $, the observable distance: $ (dl^{\cal G'})^2 =\gamma_{\alpha \beta} dx^{\alpha} dx^{\beta}, \gamma_{\alpha \beta} =-g^{\cal G '}_{\alpha \beta} + g^{\cal G'}_{0 \alpha} g^{\cal G'}_{0 \beta}/g^{\cal G'}_{00} $. 

\begin{figure}[ht]
	\begin{center}
	\includegraphics*[width=90mm]{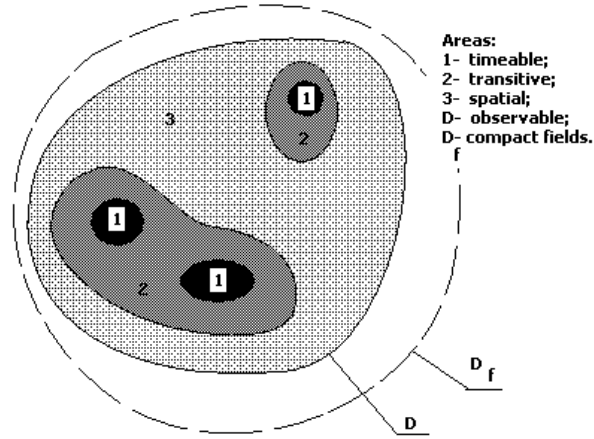}
	\caption{Observable structure of the enclosed continuum.}
	\label{fig:Struct}
	\end{center}
\end{figure}

It is clearly seen from the considered example, that the area of observation $D $ of the enclosed continuum in turn breaks up to three subareas (see Fig.\ref {fig:Struct}). In timeable areas the types of metrics of local area of an off-site continuum and the area $D $ of a continuum of the observer coincide with each other and it is possible to establish point-to-point correspondence between observable elements of off-site physical objects, therefore these off-site physical objects can be detected locally in system of the observer. The observation of off-site physical objects from spatial areas in system of the observer can be extremely unusual because of discrepancy of metrics. In some particular cases it may seem, that cause-effect chains are broken in off-site continuums. In transitive area it is possible the mixed perception of objects depending on conditions of observation.  

Generally, observable elements of metric tensor depend on all coordinates $x^i $ in system of the observer, so the internal structure of the enclosed continuum is not static for the off-site observer. Furthermore, even if the area of observation of off-site continuum $D $ coincides with all space of the observer, physical objects and their images may not coincide with each other. These conclusions were made for a special case of functional transformation (\ref{function}) corresponding to spatial rotations in works \cite{NB06, NB01}. Running forward, we shall note, that there are restrictions on possible internal structures of the enclosed continuums. In section \ref{sec:Properties} it will be shown, that the presence of the compact self-consistent fields generated by the enclosed objects in area $D_f $ (shown on Fig.\ref{fig:Struct}) is necessary. 

\section{\bf Physical objects of enclosed continuums.}
\label{sec:Properties}

The visual parameters of physical objects of off-site systems will be perceived by the observer within the framework of his own continuum. Hence, following the analysis of Sect.\ref{sec:Continuum} it will be quite logical to assume, that any physical objects (including also off-site ones) should satisfy in a continuum of the observer to the conservation laws generated by this continuum. We shall call this assumption as {\it a principle of compatibility} of off-site continuums. 

The observer registers the visual parameters of physical objects of the enclosed continuum inside some limited area $ D \subset {\cal G} $. If some given characteristic can be described by the real-valued function, one shall determine this function as $P(x^i) \equiv P(x^\alpha, \tau), i=0.. 3, \alpha=1..3 $ in some fixed point of space $x^i \in {D} $ and $P(x^i) \equiv 0$ for $ \forall x^i \not\in {D} $. Through the limitation in space, the function $P(x^\alpha, \tau) $ in each points $x^\alpha \in {D} $ can be considered as a function describing a stationary casual process and, hence, is submitted as a sum of its averaged on time value $A_0 $, some row of periodic components and the aperiodic one $p(x^\alpha, \tau) $, which averaged in time value should be equal to zero: 

\begin{eqnarray}
\fl P(x^{\alpha},\tau)= A_0(x^{\alpha})+\sum^{\infty}_{k=1}A_k(x^{\alpha} ) \cos\left[\Omega_k(x^{\alpha} ) \tau + \phi_k(x^{\alpha} ) \right] + p(x^{\alpha}, \tau) = \\ = Re\left[ \sum^{\infty}_{k=0}a_k(x^{\alpha} ) e^{i\Omega_k(x^{\alpha}) \tau} \right] + p(x^{\alpha}, \tau) = Re\left[ \sum_{\Omega}P_\Omega (x^\alpha) e^{i\Omega \tau} \right] + p(x^{\alpha}, \tau),  \nonumber 
\label{wfunc}
\end{eqnarray}

where $ \Omega_0 =0, a_0=A_0 (x^{\alpha}) , a_k = A_k (x^{\alpha}) e^{i\phi_k (x^{\alpha})} $. From the set of all frequencies $ \Omega_k (x^{\alpha}) $ on all set of observation region $ {D} $ we shall choose and fix the frequency $ \Omega $. We shall allocate the set of points $x^{\alpha}_{\Omega} $ on which the harmonic with frequency $ \Omega $ is not equal to zero. On this set we shall determine the function $P_{\Omega}(x^{\alpha}) = a_{\Omega} (x^{\alpha}_{\Omega}) $ and  determine it with zeros in other points of $ {\cal G} $.

Off-site physical objects should manifest themselves somehow as some ``influences", known fields in a system of the observer, differently it would not be any sense to examine and try to describe them. On the other hand, according to a principle of compatibility, they need to satisfy to conservation laws in the system of the observer. So, for example, they cannot be an infinite energy source in the system of the observer. There are two opportunities to satisfy to this condition: to consider, that raised fields do not transfer energy or that they are located in some limited space region in the system of the observer. 

Let's assume, that the off-site physical object described by function $P(x^\alpha, \tau) $, in system of the observer $ {\cal G} $ can be a source of the wave type field $u(x^\alpha, \tau) $, submitting to the wave equation and on each harmonic $ \Omega $ (4) should submit to the Helmholtz equation: 

\begin{equation}
\left(\frac{\partial^2 }{\partial {\tau^2}} -\frac{\partial^2 }{\partial {x^{\alpha}}^2} \right) u(x^\alpha,\tau) = P(x^\alpha,\tau), \: \:  (\nabla^2 + \Omega^2) U(x^\alpha) = - P_\Omega (x^\alpha). 	
\label{waveeq}
\end{equation}

There are partial solutions of Helmholz equation located in space. For example, in an one-dimensional case for $P_\Omega(x) = \delta(x+a) \pm \delta(x-a) $ one can get the partial solutions (even mode): $U^{in} (x) = \pm e^{\pm i\Omega a} / (i\Omega) \cos (\Omega x) $ for area $D: |x | <a $ and $U^{ex} (x) = \pm e^{\mp i\Omega x} / (i\Omega) \cos (\Omega a) $ for $ |x |> a $. At $ \Omega a =-\pi/2 + \pi n $: $U^{ex} (x) = 0 $ for $ |x |> a $, while $U^{in} (x) \not\equiv  0 $ for $ |x | <a $. 

Partial solutions of the Helmholtz equation (\ref {waveeq}) can be found in 2D- and 3D-cases. Solutions exist as in classes of continuous and discontinuous functions. As an example, in a 3D-case for $ P_\Omega (x^\alpha) = \delta(r-a) /(4 \pi a^2), $ $ r =\sqrt{x_\alpha x^\alpha} $ it is possible to get the partial solutions of three-dimensional Helmholtz equation from a class of continuous functions: $U^{in}_{even}(r) = - i e^{-i\Omega a} \cos (\Omega r) / (\Omega ar), \; U^{ex}_{even}(r) = - i e^{-i \Omega r} \cos (\Omega a) / (\Omega ar) $. Under the condition of $ \Omega a = - \pi/2 + \pi n $ fields appear located in space in region $ |r|< a $. Such conditions we shall call {\it conditions of quantization}. 

In a concrete case of localization of the electromagnetic waves excited by sources $ P(x^\alpha, \tau) \delta (S_0) $, located on a surface $S_0 $ of some area $ {D_0} \subset {D} $, sources should shield the electromagnetic waves extending inside area $D_0 $ and carrying energy. The surface $S_0 $ can be interpreted as an internal surface of the resonator. Conditions of reflection will be equivalent to conditions of equality to zero on this surface of tangential components of an electromagnetic field. For many simple configurations of resonators the distributions of electric and magnetic fields are well investigated. For each resonator, depending on its design, there is a discrete set of resonant frequencies. In our case we interpret these conditions as {\it conditions of quantization} of physical objects of the enclosed systems. 

Let's note, that electromagnetic fields, which components are normal to a surface $S_0 $, can be not limited neither this surface, nor area $D $, that does not contradicts to our principle of compatibility so as these fields do not transfer energy from area $D_0 $. In other words, only such distributions of stable off-site objects are permitted at which the fields excited by these objects in the continuum of the observer are not carrying away the energy. Fields not carrying energy can be non-local, but their own energy should be finite. 

The energy conservation law in the system of the observer can be satisfied. Necessary conditions of its realization will be {\it conditions of quantization} of observable frequencies and sizes of physical objects of off-site systems. Therefore the aperiodical component in (4) should be equal to zero and expression (4) for the enclosed off-site objects coincides with the wave function, describing quantum mechanical objects. 

In the given section the examples of solutions of the equations (\ref{waveeq}) with a source function from a class of the generalized functions \cite{Vlad81} were examined, that at all is not a necessary condition. The analysis of the generalized functions allows to find fundamental solutions and to reveal the basic properties of the common solutions. 

\begin{figure}[ht]
	\begin{center}
	\includegraphics*[width=100mm]{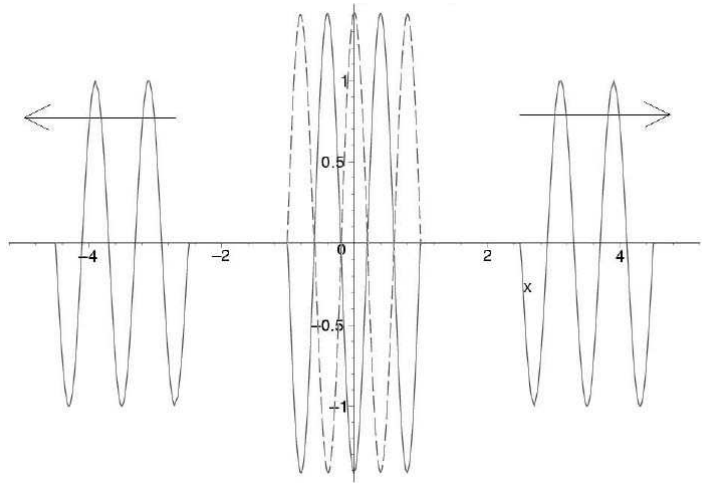}
	\caption{Occurrence of the physical object of the enclosed system.}
	\label{fig:ObjEst}
	\end{center}
\end{figure}

The Helmholtz equation gives the established solutions. The analysis of solutions of the wave equation (\ref{waveeq}) shows, that in process of the enclosed object ``occurrence" there are formed also the physical objects ``flowing out of him" at the initial time moment with speed of the excited waves. This process is shown on Fig.\ref{fig:ObjEst} at one-dimensional case for $P(x, \tau) = \left[\delta (x+1) + \delta (x-1) \right] \Theta(\tau) \sin(\Omega \tau)$, where $ \Theta(\tau) =0$ at $\tau <0$ and $\Theta(\tau) =1$ at $\tau \geq 0$. 

Thus, the off-site object can not irradiate the energy in a system of the observer, if object's spatial sizes and frequencies are quantized. The off-site object is surrounded by the compact self-consistent electromagnetic fields locally excited by him in a system of the observer. On Fig.\ref{fig:Struct} this area is shown as $D_f \supset D $, emphasizing the fact, that the fields which are not carrying energy can leave far beyond areas $D_0 $ and $D $. The occurence and the transition between stable states of the physical object is accompanied by formation of ``flowing out" objects in a system of the observer. 

The described observable properties of physical objects in micro-scale correspond to properties of quantum-mechanical and quantum-physical objects. So, from the point of view of the off-site observer, the enclosed object can have the central part like a nucleus, surrounded with the compact self-consistent electromagnetic fields excited by him. The central part, in turn, has the internal structure and in timeable areas the off-site physical objects can be observed locally in space and, so, can be identified in a system of the observer. It means, that they can participate in ``point" interactions with other objects from the system of the observer. Spatial capturing or confinement of off-site physical objects from the timeable region (area 1, Fig.\ref{fig:Struct}) in the borders of the observable region $D$ will seem as the extremely unusual phenomenon. At the standard field interpretation, it is equivalent to the presence of some force holding these objects inside the space area, and it will seem, that the quantity of this force grows at approach to the border of this area. Analogies with the phenomenon of confinement and strong interactions are arising. 

The compact self-consistent electromagnetic fields of the enclosed object will possess explicit quantum properties for the observer. Quantization is a consequence of the self-consistency, that is necessary for submitting to the energy conservation law in a system of the observer. These fields are good candidates for a role of weak interactions. Such interpretation is already confirmed experimentally as a fact of equivalence of weak and electromagnetic interactions at high energy levels.  

If all functions of transformation of continuums (like in (\ref{function})) are limited, the generalized observer should see the occurrence and disappearance in the limited area of space of some objects. From the point of view of the analysis of the given section it means, that the function of the enclosed object limited in time can be written as $P(x^\alpha,\tau)= P_0(x^\alpha,\tau) \Theta(\tau-T_0) \Theta(T_1-\tau)$, where $\left(T_0 , T_1 \right)$ is the time interval of existence of the enclosed object in system of the off-site observer. Occurrence and disappearance of physical object of off-site system will be connected to emission and absorption of quanta of the field excited by objects in system of the observer. Most likely, continuous process of an exchange in similar quantums of some system of the off-site objects enclosed and limited in time should be initiated. Generally speaking, it does not contradict to modern representations of the internal structure of physical vacuum. 

\section{\bf Physical objects of containing continuums.} 
\label{sec:MMScale}

In section \ref{sec:PhysObj} it was mentioned, that, most likely, there will be some specificity at the description of parameters of off-site containing continuums, in comparison with enclosed ones. Anyhow, physical objects of the enclosed continuums can be observed entirely while only the limited area of a containing continuum is accessible to the off-site observer from the enclosed continuum. Clearly, that the analysis of structure of an containing continuum is possible only (if it is possible in general) with the help of extrapolation of the observable data by additional physical assumptions. We shall try to receive the maximum possible information without making the additional assumptions concerning the structure of off-site continuums since the analysis of existing experimental data and possible variants can leave far beyond the framework of one article. 

Let's try to imagine, how could the observer of the enclosed continuum see an off-site continuum containing it. First of all, we shall note, that the observer of the enclosed continuum $ {\cal G} $, generally speaking, can not perceive the limitation of his own continuum on spatial and time coordinates in an off-site continuum $ {\cal G'}$ containing him ${\cal G} \rightleftharpoons D' \subset {\cal G'} $. Arguing similarly to Section \ref{sec:PhysObj} we shall come to the conclusion, that the area $D' $ from the point of view of a containing continuum can be shared into three subareas: timeable, spatial and transitive, depending on concurrence of a kind of their metrics. From other hand, their concurrence of metrics is the same for the observer of the enclosed continuum. So, there will also be the corresponding timeable, spatial and transitive areas in a containing continuum of the observer of the enclosed one. Clearly, that seen parameters of off-site physical objects of a containing continuum from the enclosed system will depend on local position of the observer in these areas in $ \cal G $, from his local metrics. It will be possible to identify locally the off-site objects of containing continuum with their images in enclosed one only from timeable areas of a continuum of the observer. 

There will be a fourth not observable area $ {\cal G'} $, distinct from $D' $ for the observer of the enclosed continuum. Not observable area will be outside the observer's continuum. However, influences from physical objects of not observable area can be registered by the observer, at least because they should influence space-time structure of area $D' $ and, hence, on seen parameters and movements of physical objects in this area. It is also possible to observe the streams of the particles emitted by physical objects of not observable area. These particles can penetrate the area $D' $, that can be perceived as the streams of particles originating by an invisible source, from ``empty" space of observer's continuum. All listed objects and processes marvelously well correspond to not clear and rather specific objects found out in last years by astrophysicists -- to ``dark matter", ``dark energy" \cite {Koch04}. Reports about the registration of intensive sources of particles from ``empty" areas of the universe have appeared also.  

In Section \ref{sec:Properties} the fields induced by objects of the enclosed continuum in the containing continuum were investigated. These fields exist in area $D' $ and also can be registered by the observer of the enclosed continuum $ {\cal G} $, for example, as the stream of particles penetrating all system of the off-site observer. These fields, as distinct from considered above, have other nature and should correlate with the sizes (or the ``age" in the standard extended Universe model) of a continuum of the observer. Observed streams of relic radiation are good candidates for this role. 

From the analysis of the previous section it is also follows, that the observable structure of our universe is (or in macro-time scale will aspire to) the structure of stable physical object in a containing continuum. Unfortunately, telling the truth, at a modern level of knowledge of astrophysical objects, it is not absolutely clear as far as last information can be useful to us. 

For completeness of a picture it is necessary to tell a few words about an opportunity of existence and observation of off-site objects in some ``average", ``human" scale, scale of the observer. Really, if physical objects of off-site systems meet constantly in micro- and macro-scales, they should be paid attention of researchers with their unusual properties if existing in ``average" scale. 

Let's consider such unusual and poorly investigated objects as fireballs. As known, they are formed as a result of the high-energy electric discharge and very much remind by external attributes to self-consistent compacts described above. It is completely not necessary, that the compact is initiated by an off-site object. It can be initiated by the system of sources of an electromagnetic field excited by the electric discharge in some environment. The fireball disappears as a result of collision with macroobject or because of gradual outflow of energy as a result of parasitic radiation because of heterogeneity of environment. When internal energy of a fireball reaches minimally possible value for maintenance of the compact, it instantly collapses with the residuary energy liberation of an electromagnetic field. And, owing to discreteness of frequencies in the compact, the character of external damages from such explosion often has strongly expressed wave character.

From Section \ref{sec:Properties} it is known, that, according to a principle of  compatibility, the necessary attribute of stable off-site object are the stable compact self-consistent fields around him. The self-consistent structure can be easily broken by more energetic external influences -- external fields or other objects as it occur in a case of a fireball. Thus, it is possible to assume, that the power density of containing system defines the scale of observation of off-site enclosed objects. In other words, the existence of the high energetic off-site objects in micro-scale interferes with their stable existence at the ``average" scale. 

\section*{\bf Conclusion.}

The hypothesis of feasible existence of off-site continuums was investigated in this paper. The opportunities of experimental observation of off-site physical objects were examined. It is appeared possible to describe the basic general observable properties of such objects even without additional assumptions of existential structure of an off-site continuum, being only based on known properties of a system of the observer. 

In a micro-scale, the scale of the elementary particles, the described properties of physical objects coincident with properties of quantum-mechanical and quantum-physical objects. Quantization is the essential property of observable parameters of physical objects of off-site systems and is a consequence of the energy conservation law in a system of the observer. There are obvious correlations with weak and strong interactions, the phenomenon of confinement. 

In a macro-scale, the scale of astrophysical objects, the appearance of off-site objects is very similar to the influence found out in last years from ``dark matter" and ``dark energy". Analogies to known relic radiation are also looked through. Such sight can appear useful at the investigation of actual problems of modern astrophysics. 

The proposed hypothesis allows unifying relativistic and quantum-physical approaches existing for today in modern physics in not contradicting system.  

\section*{\bf REFERENCES.}


\begin{thebibliography}{99}

\bibitem{Schr86} Schr\"odinger E. Space-time structure. -- Cambridge at the University press, 1950. 

\bibitem{LL88} Landau L.D., Lifshitz E.M. Theoretical Physics: The Classical Theory of Fields, vol.2 -- Pergamon Press, 1988. 

\bibitem{NB06} Novikov-Borodin A.V. Relativistic and Quantum Properties of Spatial Rotations.--  Sc.Conf. ``Physics of  Fundamental Interactions", ITEP, Moscow, Dec.2005/ LANL E-Print Archive: {\bf physics/0601155}, v.1, 20 Jan. 2006. 



\bibitem{NB01} Novikov-Borodin A.V. Space Rotation Invariance.-- DESY Report M 01-02, May, 2001/ LANL E-Print Archive: {\bf quant-ph/0105011}, May, 2001. 

\bibitem{Vlad81} Vladimirov V.S. Mathematical Physics Equations. -- Moscow: Nauka, 1981.

\bibitem{Koch04} Kochanek C.S. Where does the dark matter begin? -- arXiv:astro-ph/0412089, 2004.


\end{thebibliography}
\end{document}